# Visualization of Collaborative Data

**Guobiao Mei**
University of California, Riverside
gmei@cs.ucr.edu

**Christian R. Shelton**
University of California, Riverside
cshelton@cs.ucr.edu

## Abstract

Collaborative data consist of ratings relating two distinct sets of objects: users and items. Much of the work with such data focuses on filtering: predicting unknown ratings for pairs of users and items. In this paper we focus on the problem of visualizing the information. Given all of the ratings, our task is to embed all of the users and items as points in the same Euclidean space. We would like to place users near items that they have rated (or would rate) high, and far away from those they would give low ratings. We pose this problem as a real-valued non-linear Bayesian network and employ Markov chain Monte Carlo and expectation maximization to find an embedding. We present a metric by which to judge the quality of a visualization and compare our results to *Eigentaste*, *locally linear embedding* and *co-occurrence data embedding* on three real-world datasets.

## 1 Introduction

Collaborative data, which are composed of correlated *users* and *items*, are abundant: movie recommendations, music rankings, and book reviews, for example. Most of the work on such data has been in the area of collaborative filtering: making predictions or recommendations based on prior user ratings.

However, no previous algorithms have approached the problem of visualizing collaborative information. In this work, we initiate this problem and propose an approach. Many visualization problems are "soft" in nature and it is difficult to compare alternative methods. For this task, we introduce a simple evaluation criterion which is natural and allows for numeric comparisons of possible visualizations.

Using all the ratings, the visualization problem is to extract the intrinsic similarities or dissimilarities between all the users and items involved, and represent them graphically. This has a wide range of applications, for example guided on-line shopping. Traditional stores allow for easy browsing by physically walking up and down the aisles and visually inspecting the store's contents. Such browsing is not easy on-line. *Amazon.com*, for instance, has thousands of items in many categories. While collaborative filtering allows an on-line seller to recommend a list of objects that the buyer might also like, it does not supply a good way of browsing an on-line collection in a more free-form fashion. We propose building an embedded graph of all the items using the collaborative rating data, and allowing the shopper to zoom in on a portion of the graph and scroll around as he or she searches for items of interest. If constructed well, nearby items will also be of interest to the shopper and local directions in the space will have "meaning" to the user. Spatial layouts have been shown in the past to increase interest in exploration and to aid in finding information (Chennawasin et al., 1999).

We assume a $D$-dimensional Euclidean space, which we call the *embedded space* (for most computer interfaces, $D = 2$). Each user or item is represented by a point in this space. Intuitively, if two user (or item) points are near each other in the embedded space, the two users (items) are likely to have similar preferences (properties). In the same manner, the closer a user point is to an item point in the embedded space, the higher the rating the user has given (or would give) the item.

We formulate the visualization problem in Section 2, and then propose with our approach in Section 3. In Section 4 we show our results on three real-world datasets.

### 1.1 Prior Work

There are many existing collaborative filtering algorithms focusing on the task of prediction. We cannot review all of them and they do not directly pertain to the task of visualization. Breese et al. (1998) classifies the approaches into two major categories: *memory based* and *model based*. Memory based collaborative filtering algorithms make pre-dictions according to all the existing preferences stored beforehand, while model based algorithms first try to learn the parameters of a particular model for the existing user preferences, and then make predictions according to the learned model. Our work is mostly a model based approach.

Some of the filtering methods have a "geometric" flavor. Pennock et al. (2000) propose a collaborative filter based on personality diagnosis. They associate each user with a vector $R^{true}$, indicating the true rating of this user for every item in the system. The actual rating is assumed to be a random variable drawn from a Gaussian distribution with mean equals to the corresponding element of that item in the user's $R^{true}$ vector. If there is a missing rating, the corresponding $R^{true}$ element is a uniform random variable.

Goldberg et al. (2001) propose Eigentaste (ET). It has two phases: offline and online. It treats the entire rating matrix as a high dimensional space and employs principal component analysis for dimensionality reduction during the offline phase. It projects the users into a low dimensional space and then partitions the embedded space into sets of users and uses the maximally rated items in a given set as predictions during the online phase. While this algorithm does place the users in a geometric space, it does not place the items in the same space, and it requires that there be a set of items (the *gauge set*) which every user has rated.

Saul and Roweis (2003) propose locally linear embedding (LLE) as an algorithm for embedding points into a (low dimensional) space. It is not geared to collaborative data: it does not deal well with sparse data, and it can only embed either users or items. However, it has had good success with non-linear embeddings, so it is a candidate method for the collaborative visualization problem.

Co-occurrence data have been used to produce embeddings of two classes of objects in the same space. CODE (Globerson et al., 2005) is one such recent example. It tries to embed objects of two types into the same Euclidean space based on their co-occurrence statistics. Unfortunately, collaborative filtering data are usually given as ratings and not co-occurrence statistics. Even if we take the ratings to be proportional to the corresponding co-occurrences (an unjustified assumption), we still have missing statistics (which cannot be taken to be zero). Co-occurrence algorithms do not currently deal with such missing values.

None of the above three algorithms (ET, LLE and CODE) was developed for our task. However, there are no other algorithms designed for collaborative visualization. Therefore, we compare to them (to the extent possible) in Section 4.

## 2 The Visualization Problem

We first introduce our notations in 2.1. In 2.2 we formulate the visualization problem of collaborative data. We specify our distributional assumptions in 2.3.

### 2.1 Notation

Let $U = \{u_1, u_2, \ldots, u_m\}$ and $G = \{g_1, g_2, \ldots, g_n\}$ be the sets of all users and items, respectively. Without ambiguity, we will use $u_i$ to refer both to the $i$-th user and to the corresponding point in the embedded space for that user. We use the same notation for $g_j$. We define $\delta_{ij}$ to be 1 if $u_i$ rated $g_j$ and 0 otherwise.

Let $r_{ij}$ be the rating of $u_i$ of $g_j$ if $\delta_{ij} = 1$. Here we normalize all the ratings to the range $[0, 1]$ (*i.e.* 1 is the highest rating, and 0 is the lowest). Let $R = \{r_{ij} \mid \delta_{ij} = 1\}$.

We further denote $G^i = \{g_j \mid \delta_{ij} = 1\}$ and $U^j = \{u_i \mid \delta_{ij} = 1\}$. $G^i$ is the set of all the items that $u_i$ rated, and $U^j$ is the set of all the users that rated $g_j$.

### 2.2 Formulation of the Problem

The visualization problem is to find an embedding of all the points $U$ and $G$ in a Euclidean space we call the *embedded space*. The embedding should be one in which the distance between a user and an item is related to the corresponding rating.

In the embedded space, we assume each $u_i$ and $g_j$ are random variables drawn independently from prior distributions $P_u(u_i)$ and $P_g(g_j)$. We introduce a rating function $f : \Re_0^+ \mapsto [0, 1]$, which maps the distance between two points (a user and an item) in the embedded space to a real value on $[0, 1]$: the expected rating for the two points. $f(x)$ is a monotonically non-increasing function, $f(0) = 1$, and $f(\infty) = 0$. Intuitively, two points with a smaller mutual distance should have a higher expected rating. At this point, we will assume that the rating function $f(x)$ is given. In Section 3.4 we will show how this function can be learned from data. The actual rating $r_{ij}$ between two points $u_i$ and $g_j$ in the embedded space is a random variable drawn from a distribution $P_f(r_{ij} \mid u_i, g_j)$ with mean $f(\|u_i - g_j\|)$.

Given all the ratings, $R$, as evidence and the rating function, $f$, our task is to put all the user points $U$ and item points $G$ into the embedded space so that the likelihood of observed ratings $R$ is maximized. That is, we want to find the $U$ and $G$ points that maximize the posterior:

$$[U^*, G^*] = \arg\max_{U,G} P(U, G \mid R) \qquad (1)$$
$$= \arg\max_{U,G} \prod_{i,j \mid \delta_{ij}=1} P_f(r_{ij}|u_i, g_j) \prod_i P_u(u_i) \prod_j P_g(g_j) \ .$$

$P(U, G, R)$ is a real-valued Bayesian network in which each user and item variable has no parents and each rating variable has two parents (one user and one item). The ratings are given as evidence and the task is to determine the most probable joint assignment to the user and item variables given the ratings.

### 2.3 Gaussian Assumptions

We assume that all the distributions are from the Gaussian family. To be specific,

$$P_u = \mathcal{N}(\mathbf{0}, \mathbf{\Sigma_u})$$
$$P_g = \mathcal{N}(\mathbf{0}, \mathbf{\Sigma_g})$$
$$P_f(r_{ij}|u_i, g_j) = \mathcal{N}(f(\|u_i - g_j\|), \sigma_r) \ .$$

Note that while these distributions are all normal, the function $f$ is non-linear and therefore the resulting joint distribution is not Gaussian.

## 3 Approach

It is intractable to compute the posterior in Equation 1 directly. We use Markov chain Monte Carlo sampling.

### 3.1 Metropolis-Hastings Algorithm

In particular, we use the Metropolis-Hastings (MH) algorithm (Metropolis et al., 1953), which was extended to graphical models. Given a graphical model over the random variables $X = \{x_1, x_2, \ldots, x_N\}$, assume a target distribution $\pi$ over $X$. For each variable $x_i$, there is an associated proposal distribution $Q_{x_i}$, the distribution of new samples for that variable.

Given a current assignment to $X$, MH randomly picks a variable $x_i$ and tries to replace its value with a new sample $x_i'$ drawn from the proposal $Q_{x_i}$. Let $Y = X - \{x_i\}$.

The *transition gain ratio* for changing the sample $x_i$ to $x_i'$ is defined as

$$\mathcal{T}^Q(x_i \to x_i') = \frac{Q_{x_i'}(x_i)\pi(Y, x_i')}{Q_{x_i}(x_i')\pi(Y, x_i)} \ . \quad (2)$$

The probability of accepting this new sample $x_i'$ is

$$\mathcal{A}(x_i \to x_i') = \min\left\{1, \mathcal{T}^Q(x_i \to x_i')\right\} \ . \quad (3)$$

Using the local independencies of the graph, this can be decomposed into a set of local probabilities.

### 3.2 Sampling Approach

In our visualization problem, $\pi$ is the posterior distribution of $U$ and $G$ given $R$ (Equation 1). Initially, we sample from $P_u$ for every $u_i$ and sample from $P_g$ for every $g_j$. This jointly form a single starting sample (a joint assignment to $U$ and $G$) for our MCMC method.

We use proposal distributions $Q_{u_i}$ for the node $u_i$ and $Q_{g_j}$ for the node $g_j$. We set the proposal distributions to be Gaussian with means at the previous embedded position:

$$Q_{u_i} = \mathcal{N}(u_i, \mathbf{\Sigma_u'})$$
$$Q_{g_j} = \mathcal{N}(g_j, \mathbf{\Sigma_g'}) \ .$$

If we choose the node $u_i$ to be resampled, we draw $u_i'$ from $Q_{u_i}$, and then compute the accept ratio for this change according to Equation 3. Using the local independence properties, the transition gain ratio with respect to the rating function $f$ is given by

$$\mathcal{T}_f^Q(u_i \to u_i') = \frac{Q_{u_i'}(u_i) P_u(u_i') \prod_{j \in G^i} P_f(r_{ij}|u_i', g_j)}{Q_{u_i}(u_i') P_u(u_i) \prod_{j \in G^i} P_f(r_{ij}|u_i, g_j)} \ . \quad (4)$$

Similarly, the transition gain ratio for an item node, $g_j$ is

$$\mathcal{T}_f^Q(g_j \to g_j') = \frac{Q_{g_j'}(g_j) P_g(g_j') \prod_{i \in U^j} P_f(r_{ij}|u_i, g_j')}{Q_{g_j}(g_j') P_g(g_j) \prod_{i \in U^j} P_f(r_{ij}|u_i, g_j)} \ . \quad (5)$$

We repeat the above resampling phase until the process has mixed. The stationary distribution of this procedure is the true posterior $P(U, G|R)$.

### 3.3 Simulated Annealing

Recall that we want the $\arg\max_{U,G} P(U, G|R)$. The Metropolis-Hastings algorithm will give us joint samples of $U$ and $G$, drawn from that posterior. To get the samples that maximize the posterior, we modify the standard MH algorithm along the lines of the simulated annealing (SA) algorithm (Kirkpatrick et al., 1983).

In particular, we modify Equation 2 to add an annealing temperature, $\beta$:

$$\mathcal{T}^Q(x_i \to x_i') = \frac{Q_{x_i'}(x_i)\pi(Y, x_i')^\beta}{Q_{x_i}(x_i')\pi(Y, x_i)^\beta}$$

The transition gain ratios of equations 4 and 5 are then

$$\mathcal{T}_f^Q(u_i \to u_i') = \frac{Q_{u_i'}(u_i) \left[P_u(u_i') \prod_{j \in G^i} P_f(r_{ij}|u_i', g_j)\right]^\beta}{Q_{u_i}(u_i') \left[P_u(u_i) \prod_{j \in G^i} P_f(r_{ij}|u_i, g_j)\right]^\beta}$$

$$\mathcal{T}_f^Q(g_j \to g_j') = \frac{Q_{g_j'}(g_j) \left[P_g(g_j') \prod_{i \in U^j} P_f(r_{ij}|u_i, g_j')\right]^\beta}{Q_{g_j}(g_j') \left[P_g(g_j) \prod_{i \in U^j} P_f(r_{ij}|u_i, g_j)\right]^\beta}$$

The added temperature factor $\beta$ grows gradually from 1 to $\infty$. Initially $\beta = 1$, and this method is the same as the standard Metropolis-Hastings algorithm. As $\beta$ grows, the simulated annealing algorithm penalizes changes resulting in lower likelihoods; the algorithm tends to only climb uphill in the posterior distribution.

### 3.4 Learn the Rating Function

Until now, we have assumed that the rating function $f$ was known. However, we would like this function to adapt to the collaborative data.

It would be straight-forward to select $f$ from a family (for example the exponential, $f(x) = e^{-\lambda x}$). However, the actual rating function may have a very different shape. Instead we note that all collaborative datasets of which we are aware have a finite number of values for the ratings. Many are binary ("like" or "do not like") and others are based on a five- or ten-point scale. Continuous, real-valued ratings are seldom used. We therefore let $f$ be a step function with discrete quantizations.

We discretize $f$ into $K$ quantizations. Let $\Theta = \{\theta_i \mid i = 1, \ldots, K\}$ be the set of $K$ splitting points in sorted order, with $\theta_K = \infty$. Given the set of splitting points $\Theta$, the rating function is

$$f(x; \Theta) = 1 - \frac{i-1}{K-1}, \quad \text{if } \theta_{i-1} \leq x < \theta_i .$$

The problem of learning $f$ is now transformed to the problem of learning $\Theta$:

$$\Theta^* = \arg\min_{\Theta} \sum_{i,j \mid \delta_{ij}=1} E[(f(\|u_i - g_j\|; \Theta) - r_{ij})^2] \quad (6)$$

where the expectation is with respect to the posterior distribution over $U$ and $G$. This formulation is equivalent to maximizing the probability of the ratings; the squared error in the above equation comes directly from the Gaussian assumption regarding the distribution $P_f$.

We use expectation maximization (EM) algorithm (Dempster et al., 1977) to learn the rating function. We initially set $\Theta = \Theta^0$, a random starting point that meets our requirements for $f$.

The E-step employs MH to sample from the expectations in Equation 6 using the rating function $f^k = f(\cdot; \Theta^k)$ at the $k$-th iteration.

The M-step updates $\Theta^{k+1}$ based on the generated sample configurations of the embedded space (which approximate the expectations of Equation 6). Using all the $u_i$ and $g_j$ points, the optimal rating function is updated according to Equation 6. Let $N$ be the number of terms in the summation of Equation 6 (one for each rating for each sample). The M-step optimization can be done efficiently (and exactly) in $O(NK)$ time using dynamic programming.

Before updating the rating function in the M-step, we renormalize all the points in the embedding space. Due to our assumptions that $P_u$ and $P_g$ are fixed Gaussian distributions with zero mean and that we have the freedom to change $f$, if the above procedure were run without modification, all the points would collapse together toward the origin. Consequently, the learned rating function would have splitting points with smaller and smaller values. We fix this by a simple normalization step that scales and translates the points to reset the mean of all of the points to zero and the variance of their positions to one. Note that this is not a general "whitening" step in that we only multiply the points by a scalar, not a matrix.

---

**Algorithm 1** $[U, G] \Leftarrow$ Embed-Graph($R$,$D$)
  **Inputs:** $R$: rating matrix, $D$: embedding dimensionality
  **Outputs:** $U$ and $G$: embedded points

  $\beta \Leftarrow 1$
  $P_u \Leftarrow \mathcal{N}(\mathbf{0}, \mathbf{\Sigma_u}), P_g \Leftarrow \mathcal{N}(\mathbf{0}, \mathbf{\Sigma_g})$
  $Q_{u_i} \Leftarrow \mathcal{N}(u_i, \mathbf{\Sigma'_u}), Q_{g_j} \Leftarrow \mathcal{N}(g_j, \mathbf{\Sigma'_g})$
  $P_f(r_{ij} \mid u_i, g_j) \Leftarrow \mathcal{N}(f(\|u_i - g_j\|), \sigma_r)$
  $f \Leftarrow f(\cdot; \Theta^0)$
  Sample $\{u_i \sim P_u\}_{i=1}^m$
  Sample $\{g_j \sim P_g\}_{j=1}^n$
  **repeat**
      // *E-Step:*
    $\mathcal{S} \Leftarrow \emptyset$
    **for** $k = 1$ to $l_b + l_s$ **do**
      Randomly pick a point $x_i$ from samples in $[U, G]$
      **if** $x_i$ is a user point $u_i$ **then**
        Sample $u'_i \sim Q_{u_i}$
        $u_i \Leftarrow u'_i$ with probability $\mathcal{A}_f(u_i \to u'_i)$
      **else if** $x_i$ is an item point $g_j$ **then**
        Sample $g'_j \sim Q_{g_j}$
        $g_j \Leftarrow g'_j$ with probability $\mathcal{A}_f(g_j \to g'_j)$
      **end if**
      **if** $k > l_b$ **then**     // *burn in for $l_b$ iterations*
        Add $(U, G)$ to $\mathcal{S}$     // *Save last $l_s$ iterations*
      **end if**
      $[U, G] \Leftarrow$ normalize($[U, G]$)
    **end for**
    $\beta \Leftarrow (1 + \epsilon)\beta$
      // *M-Step:*
    $f(\cdot; \Theta) \Leftarrow$ learned rating function using $\mathcal{S}$
  **until** The current sample $[U, G]$ is stable

---

### 3.5 Overview of the Full Algorithm

To put everything together, the overall algorithm in our approach is listed in Algorithm 1. The parameters of this algorithm are $l_s$ (the number of samples used for estimating the expectation), $l_b$ (the number of samples necessary for the MCMC process to converge), $\epsilon$ (the amount by which to increase $\beta$), and the variances of the Gaussian distribu-tions. In Section 4.2 we specify these quantities for the experiments we ran.

## 4 Experiments and Results

We discuss our three datasets, our methodology for comparison, and then compare our algorithm to three others.

### 4.1 Experiment Datasets

The SAT dataset contains SAT II subject examination scores for 40 questions chosen from a study guide of historic questions and 296 users. SAT II is a standard exam taken by high school seniors applying to colleges in the United States. All the scores are either 0 or 1 (indicating whether the student got the question correct), and there are no missing values. The 40 questions are from the subjects French, Mathematics, History and Biology. The exam was administered on-line over the course of one week.

The BGG dataset comes from *www.boardgamegeek.com* which contains ratings from thousands of game players and thousands of board games. The ratings range, in half-integer increments, from 0 to 10. We picked the 400 users and 80 games with the highest rating density. The rating matrix density, which we define as $\frac{\sum_{i,j} \delta_{ij}}{\sum_{i,j} 1}$, is 63.4% for this subset. Our snapshot of the dataset is from January 2005. The ratings are available publicly from the website.

Finally, the MovieLens dataset contains ratings from users on a variety of movies. All the ratings are integers from 1 to 5. We picked 400 users and 50 movies, again to maximize the rating density. The rating matrix density is 41.0% on this subset. This dataset is publicly available from *movielens.umn.edu*.

### 4.2 Algorithm Initialization

Because we are learning the rating function $f$, the absolute positions of the embedded points will not affect our approach directly. Rather, the relative positions of each point matter. Therefore, the overall scale of $\Sigma_\mathbf{u}$ and $\Sigma_\mathbf{g}$ do not affect the result.

In particular, for all the three datasets, we set $\Sigma_\mathbf{u}, \Sigma_\mathbf{g}, \Sigma'_\mathbf{u}$ and $\Sigma'_\mathbf{g}$ each to be the identity matrix. We set $\sigma_r = 0.25$ for SAT, $\sigma_r = 0.1$ for MovieLens, and $\sigma_r = 0.05$ for BGG. These values directly reflect the discretization of the rating scores. Our informal tests show that the algorithm is not sensitive to these particular numbers and we have made no effort to tune them.

For the rating function $f$, we choose to set $\Theta^0$ directly using an M-step from the samples drawn from their priors. For our experiments, we set $l_s = 2000$, $l_b = 1000$, and $\epsilon = 0.02$ for all three datasets.

### 4.3 Implementation Issues

We compare our results with *Locally Linear Embedding* (LLE) (Saul & Roweis, 2003), *Eigentaste* (ET) (Goldberg et al., 2001) and *co-occurrence data embedding* (CODE) (Globerson et al., 2005). None of the algorithms is exactly suited to our problem, so we discuss our adaptations in this section.

If we consider the rating matrix as a set of points in the high dimensional space, we can use LLE to embed them into a lower dimensional space. The LLE algorithm requires a full rating matrix $R$. This is not available for the MovieLens and BGG datasets. We use linear regression to fill the missing ratings. To be specific, we first fill each missing rating $r_{ij}$ with average rating of $u_i$. This results in full rating matrix. To predict the missing rating $r_{ij}$ using linear regression, for each item $g_k$, let $\mathbf{x}_k = [r_{1k}, \ldots, r_{(i-1)k}, r_{(i+1)k}, \ldots, r_{mk}]^\top$, *i.e.* $\mathbf{x}_k$ is the vector containing all the ratings for $g_k$ except from $u_i$. We then use linear regression to find

$$\left[\hat{\mathbf{w}}_i, \hat{b}_i\right] = \arg\min_{\mathbf{w},b} \sum_k (\mathbf{w}^\top \mathbf{x}_k + b - r_{ik})^2$$

The predicted rating is then given by

$$\hat{r}_{ij} = \hat{\mathbf{w}}_i^\top \mathbf{x}_j + \hat{b}_i \ .$$

Both LLE and ET can embed either users or items into an Euclidean space. Yet, neither of them can embed both in the same space. We tried several ways to extend them and to make them comparable. One straight-forward way is to embed all the user points first into the space. Then for every item, find all the users who gave it its highest rating, and place this item at the mean of those users points.

For our results, we used an alternative method, which performed better than the one above. Let $\hat{R}$ be the full rating matrix filled in using linear regression. We introduce a correlation matrix $C$ among all $n$ items. The diagonal $C_{ii}$ is set to 1. Let $R_i$ be the $i$-th column of $\hat{R}$,

$$C_{ij} = \frac{R_i^\top R_j}{\|R_i\| \cdot \|R_j\|} \ .$$

We then let $X = [C \ \hat{R}^\top]$ and use LLE or ET to embed $X$ into the target Euclidean space. The first $n$ points correspond to the items and the last $m$ points to the users.

ET only works if there is a *gauge set* of items which all users have rated. However, in the MovieLens and BGG datasets, no such gauge set exists. Using the above regression technique to fill in a gauge set results in bad (and misleading) results, so we omitted them and have only included ET results for the SAT dataset.

The CODE algorithm requires co-occurrence statistics between users and items. The relationship between co-occurrences and ratings is not clear. However, it is natural to assume $r_{ij}$ is proportional to the probability of the co-occurrence of $u_i$ and $g_j$. Intuitively, a higher rating indicates it is more likely that the user and item "occur" at the same time. We set the empirical distribution of $(u, g)$ to be proportional to the rating matrix (filled by linear regression if there are missing ratings). We initialize the mappings uniformly and randomly from the set $[-0.5, 0.5]^D$ as the starting point for the optimization.

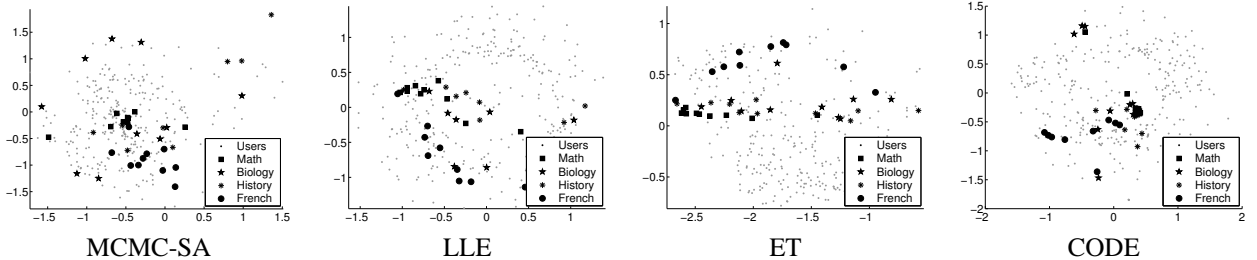

Figure 1: 2-Dimensional embeddings for the SAT questions using a simulated annealing version of the MCMC algorithm, local linear embedding, Eigentaste, and CODE.

Our linear regression method for filling in missing values has proven reasonable on the prediction task, but admittedly it is not the most sophisticated algorithm possible. Therefore, to distinguish embedding factors from data completion factors, we also ran our MCMC algorithm on the completed rating matrix from linear regression.

Both our method and CODE have variable running times (number of EM iterations in our case, number of random restarts for CODE). For the results reported here, we gave each 30 seconds of CPU time on a 2.8 GHz processor.

### 4.4 Sample Results

The SAT data was selected because of our ability to extract a "ground truth." In particular, we expect that when embedded, the questions from the same subjects should be grouped together. Figure 1 shows the embedding for dimension 2 using the simulated annealing approach (along with the three other approaches).

There are ten questions in each category. We can clearly see that our method clusters all the French questions tightly together. The same happens for the Math questions. (There are eight Math questions that overlap in a small area.) The other methods do not produce as tight clusters.

The History and Biology questions do not cluster as well. Further data analysis has shown that there is very little predictability in the History and Biology questions, so this result is perhaps not surprising.[1]

---
[1]The French and Math questions tended to test a body of knowledge that is often retained as a coherent block, where as the History and Biology questions on this exam tended to test more isolated blocks of knowledge.

Figure 1 also shows that the user points and the item points intermix more evenly with our approach. This meets our expectation that for any user, we can always find things they like or dislike (questions on which they perform well or poorly). In the embedding results of LLE, ET, and CODE, a large number of user points lie in parts of the graph outside the convex hull of the the question points. This makes it impractical to make recommendations to those users based on the visualizations.

### 4.5 Evaluation Criteria

There are no prior standard metrics for evaluating the quality of the embedded graph. In this work, we introduce *Kendall's tau* (Kendall, 1955) as a suitable evaluation criterion. Kendall's tau is used to compute the correlation in ordering between two sequences $X$ and $Y$. It is especially useful for evaluating the correlation between two sequences that may have many ties.

Given two sequences $X$ and $Y$ of the same length, a pair $(i, j), i \neq j$ is called *concordant* if the ordering of $X_i$ and $X_j$ is the same as the ordering of $Y_i$ and $Y_j$. By contrast, if the relative ordering is different, this pair $(i, j)$ is called *discordant*. If $X_i = X_j$ or $Y_i = Y_j$, then $(i, j)$ is neither concordant or discordant, and it is called an *extra x* pair or *extra y* pair, respectively.

Kendall's tau is defined as

$$\tau = \frac{C - D}{\sqrt{C + D + E_y}\sqrt{C + D + E_x}} \qquad (7)$$

where $C$ is the number of all concordant pairs, and $D$ is the number of all discordant pairs. $E_x$ and $E_y$ are the numbers of extra $x$ pairs and extra $y$ pairs.

It is easy to verify that $\tau$ is always between $-1$ and $1$. $\tau = 1$ indicates the two sequences have perfect positive correlation, and $\tau = -1$ indicates perfect negative correlation. $\tau = 0$ indicates their orderings are independent.

To evaluate the quality of the graph in the embedded space, for each test we randomly select a set of users, $\bar{U}$, and items, $\bar{G}$, as testing users and items. All the ratings between users in $\bar{U}$ and items in $\bar{G}$ are held out for testing and are not used in generating the embedding.

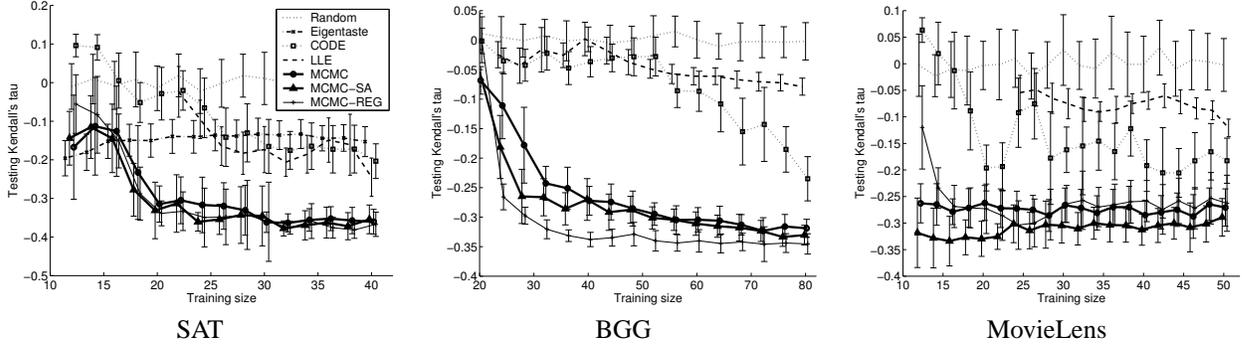

| SAT | BGG | MovieLens |

Figure 2: Performance of embedding algorithms on the three datasets as a function of training set size.

The embedding algorithms will produce the embedded points for those nodes in $\bar{U}$ and $\bar{G}$ given some additional ratings. In order to evaluate the embedding quality, we generate two sequences and compute Kendall's tau between them: sequence $X$ contains the actual ratings between all the pairs $u_i \in \bar{U}$ and $g_j \in \bar{G}$ such that $\delta_{ij} = 1$, and sequence $Y$ contains the distances between the corresponding $u_i$ and $g_j$ in the embedded graph.

A good embedding will place $u_i$ far from $g_j$ if $r_{ij}$ is small, and close if $r_{ij}$ is large. Kendall's tau for the above two sequences exactly evaluate this correlation. Denote $\tau$ to be the Kendall's tau for the sequences $X$ and $Y$. Negative values of $\tau$ indicate good embeddings, and we expect the values from embedding algorithms to be smaller than 0 (that of a random embedding).

### 4.6 Experimental Results

We ran our MCMC algorithm both with and without simulated annealing, along with LLE, ET and CODE. We randomly selected one quarter of the users and one quarter of the items for testing ($\bar{U}$ and $\bar{G}$ from above). We randomly selected other users and items to form a training set ($\tilde{U}$ and $\tilde{G}$). All ratings between members of $\tilde{G}$ and $\tilde{U}$, $\tilde{G}$ and $\bar{U}$, and $\bar{G}$ and $\tilde{U}$ are used for training. As stated previously, the ratings between members of $\bar{G}$ and $\bar{U}$ are used for testing. It is necessary to include the ratings between $\tilde{G}$ and $\bar{U}$ (and likewise between $\bar{G}$ and $\tilde{U}$) in order to connect the test users and items with the training users and items.

Because the existence of the test set, there are always missing ratings in the rating matrices used. We use linear regression to fill those ratings. We also ran the MCMC algorithm on the same filled data as LLE, ET and CODE used (MCMC-REG in the graphs).

For each dataset size (number of items in $\tilde{G}$), we ran 25 independent experiments and recorded the means and standard deviations across the experiments for all algorithms. Every algorithm was run on the same set of training and testing sets.

For each of these datasets, Figure 2 shows the comparison of our methods to LLE, ET, CODE, and a random embedding, as a function of the size of the training set. We also computed an "ideal embedding" value for $\tau$. Because of ties, Kendall's tau cannot always reach $-1$, so we calculate the lowest possible value for $\tau$ on the random dataset drawn. This takes nothing into account except ties and it is *highly* optimistic and probably not obtainable at such low dimensions. The optimal values for SAT, BGG, and MovieLens datasets are approximately $-0.63$, $-0.87$ and $-0.90$ respectively. We ran LLE algorithm with training size starting at 22 for SAT and 24 for MovieLens because of matrix inversion problems for smaller training sizes.

Figure 3 shows another experiment with the same evaluation criteria. In this experiment, we fix the testing data as usual, and use all the remaining data for training. The plot shows Kendall's tau as a function of the number of dimensions of the embedding space.

### 4.7 Analysis of the Results

From the experiments above, we can see that when the rating matrix is denser, the embedding algorithm achieve better results. Our sampling method, with (MCMC-SA) and without (MCMC) simulated annealing, outperformed LLE, ET and CODE. None of them were designed with this type of data in mind, so we do not present these results to disparage those methods, but there were no other methods available to test against. Note that our MCMC algorithm on linear regression filled data (MCMC-REG) has similar performance to directly applying our MCMC method on data with missing ratings. This implies that it is not our regression that is causing the poor results from the other algorithms, but rather their misfit to this problem. We would also note that our method also seems more stable (smaller variance) than the other algorithms compared.

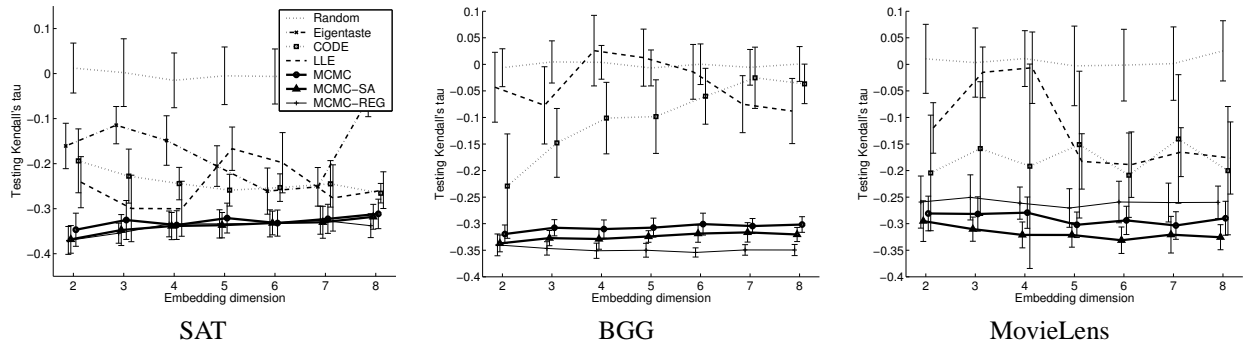

Figure 3: Performance of embedding algorithms on the three datasets as a function of embedding dimensions.

On the SAT dataset, which contains full density of ratings, our algorithms show strong negative correlations which indicate good visualization results. In most cases, using simulated annealing helps improve the quality of embedding (compared to "normal" MCMC). As the training size grows, we have more information on the relations between all the user and item points, and that leads to better performance for all the algorithms.

The BGG and MovieLens datasets have many missing ratings and the ratings values are more subjective and therefore noisier. Our algorithm is not as competitive with the "ideal" value for Kendall's tau, but we feel that this ideal value is wildly optimistic in these settings. Our algorithm does perform better than random embeddings, LLE, and CODE.

## 5 Conclusions

We formulated a new problem of visualizing collaborative data. This is a potentially very useful problem. Not only are on-line databases of user ratings growing, but personal databases are also becoming more common. We expect the collaborative visualization problem to be useful in organizing personal music or photography collections as well as on-line shopping.

We have not addressed the computational issues nor the stability of the resulting embedding in this work. Both are important problems for on-line deployment in changing databases. Because of the anytime nature of sampling methods and the ease of introducing constraints, we are hopeful that the solution presented here can be adapted to provide stable and adaptive solutions.

## Acknowledgments

We thank Titus Winters for collecting and sharing the SAT dataset. This work was supported in part by a grant from Intel Research and the UC MICRO program.